\documentclass[twocolumn,aps,superscriptaddress,showpacs,floatfix,prc,noshowpacs]{revtex4-2}
\usepackage{mathrsfs}
\usepackage{amssymb}
\usepackage{amsmath}
\usepackage{graphicx}
\usepackage[normalem]{ulem}
\usepackage[dvips]{color}
\usepackage{bm}
\usepackage{longtable}
\usepackage{slashed}
\usepackage{enumitem}
\usepackage{empheq}

\usepackage{titlesec}
\renewcommand{\thesection}{\Roman{section}}
\titleformat{\section}{\small\bfseries\centering}{\thesection.}{0.5em}{}

\usepackage{times}
\usepackage{fouriernc}
\usepackage[T1]{fontenc}
\usepackage{utopia}

\usepackage[bb=ams, cal=cm, scr=boondox, frak=euler]{mathalpha}

\usepackage[breaklinks=true]{hyperref}
\hypersetup{
  colorlinks=true,
  citecolor=magenta,
  linkcolor=black,
  urlcolor=magenta
}

\renewcommand\sout{\bgroup \color{red} \ULdepth=-.5ex \ULset}

\renewcommand{\rm}[1]{\textrm{#1}}
\renewcommand{\d}{\mathrm{d}}

\usepackage{tikz,xcolor,hyperref}

\definecolor{lime}{HTML}{A6CE39}
\DeclareRobustCommand{\orcidicon}{
	\begin{tikzpicture}
	\draw[lime, fill=lime] (0,0) 
	circle [radius=0.16] 
	node[white] {{\fontfamily{qag}\selectfont \tiny ID}};
	\draw[white, fill=white] (-0.0625,0.095) 
	circle [radius=0.007];
	\end{tikzpicture}
	\hspace{-2mm}
}
\foreach \x in {A, ..., Z}{%
	\expandafter\xdef\csname orcid\x\endcsname{\noexpand\href{https://orcid.org/\csname orcidauthor\x\endcsname}{\noexpand\orcidicon}}
}

\begin{document}

\title{Revisiting the Possibility of a Sharp Phase Transition in Cold Neutron Stars}

\author{Bao-Jun Cai\orcidA{}}\email{bjcai@fudan.edu.cn}
\affiliation{Key Laboratory of Nuclear Physics and Ion-beam Application (MOE), Institute of Modern Physics, Fudan University, Shanghai 200433, China} 
\affiliation{Shanghai Research Center for Theoretical Nuclear Physics, NSFC and Fudan University, Shanghai 200438, China}
\author{Bao-An Li\orcidB{}}\email{Bao-An.Li$@$etamu.edu}
\affiliation{Department of Physics and Astronomy, East Texas A\&M University, Commerce, TX 75429-3011, USA}
\author{Yu-Gang Ma\orcidC{}}\email{mayugang$@$fudan.edu.cn}
\affiliation{Key Laboratory of Nuclear Physics and Ion-beam Application (MOE), Institute of Modern Physics, Fudan University, Shanghai 200433, China} 
\affiliation{Shanghai Research Center for Theoretical Nuclear Physics, NSFC and Fudan University, Shanghai 200438, China}

\date{\today}

\newcommand{\x}{\mathrm{X}}
\newcommand{\y}{\mathrm{Y}}
\newcommand{\hr}{\widehat{r}}
\newcommand{\hP}{\widehat{P}}
\newcommand{\heps}{\widehat{\varepsilon}}
\newcommand{\hrho}{\widehat{\rho}}

\begin{abstract}
First-order phase transitions (FOPTs) in cold neutron stars (NSs) have been extensively studied and have provided valuable insights into the behavior of the densest matter visible in our Universe, although a strong consensus has yet to emerge. Revisiting the possibility of a hadron-quark FOPT from a new perspective, we examine the interplay between the coupled nature 
of gravity and microscopic interactions in Tolman--Oppenheimer--Volkoff (TOV) equations and the fundamental requirements of thermodynamic consistency in NSs. We demonstrate that a sharp FOPT manifested as a plateau in the equation of state (EOS) $P(\varepsilon)$, i.e., pressure $P$ versus energy density $\varepsilon$, is intrinsically incompatible with the regularity conditions of the TOV solutions. Although numerical integrations of the TOV equations with EOSs incorporating FOPTs may yield seemingly reasonable mass-radius relations consistent with current observations, such results can mask underlying inconsistencies. Our analysis thus establishes a structural consistency criterion for constraining dense-matter EOSs using NS observables, complementing existing studies of possible phase transitions in NS interiors.
\end{abstract}

\pacs{21.65.-f, 21.30.Fe, 24.10.Jv}
\maketitle

\section{Introduction}

Neutron stars (NSs) provide a unique laboratory to probe matter at extreme densities\,\cite{Shapiro1983,Walecka1974,Collins1975,Chin1976,Freedman1977,Akmal1998,LP01,LP07,Alford2008,LCK08,LCXZ21,Latt21}. The macroscopic structure of a cold NS is fully determined by its Equation of State (EOS), which relates the pressure $P$ to the energy density $\varepsilon$, namely $P = P(\varepsilon)$, at zero temperature.  A key property of the NS matter EOS is the squared speed of sound (SSS),
\begin{equation}
s^2\equiv\frac{\d P}{\d \varepsilon},
\end{equation}
which characterizes the stiffness of dense matter and the propagation of pressure perturbations\,\cite{Landau1987}.

A particularly interesting question is whether matter in NSs can undergo sharp first-order phase transitions (FOPTs) signaled by an abrupt vanishing of the SSS at supranuclear densities, i.e.,
\begin{equation}
\boxed{
s^2=0\leftrightarrow\rm{FOPT}.}
\end{equation}
A FOPT is characterized by discontinuities in $\varepsilon(P)$ as a function of pressure; that is, at a fixed $P$, two thermodynamically stable phases with different energy densities can coexist, producing a plateau in the EOS $P(\varepsilon)$. In this region, the pressure remains constant because the Gibbs free energies (or chemical potentials) of the two phases are equal, while the system converts one phase into the other as $\varepsilon$ changes continuously with the phase fraction.
Such transitions can lead to distinctive features in the NS mass-radius (M-R) relation, including cusps or twin-star configurations\,\cite{AHP13,Alford17,Xie21,Tan22-a,Jim24,Komo24,ZLi25,Grun25-a,LBA25,Sun23,ZLi23,CZhang23,LJJ24,Essick24,Pat25,Pat25-a,Pal25,HuangXR25}. Despite their theoretical appeal, the existence of sharp FOPTs in NSs remains largely uncertain with little observational evidence\,\cite{Tews18,Ferr20,Miao20,Pang20,Tang21,Tang21-a,Som23,Gorda23,Ecker23,Brad23,Tak23,Lin24,Huang25}. Within an actual NS, where gravity plays a central role,  both the pressure $P(r)$ and the energy density $\varepsilon(r)$ decrease continuously with the radial coordinate, as required by hydrostatic equilibrium and mechanical stability, even across a phase-transition region.
Based on recent astrophysical observations of NS masses, radii as well as tidal deformabilities\,\cite{Abbott2017,Abbott2018,Abbott2020-a,Riley19,Miller19,Fon21,Riley21,Miller21,Salmi22,Choud24,Reardon24,Shirke25}, a few theoretical studies suggest that a crossover may be physically more plausible\,\cite{Masu13,Huang22,Kedi22,Baym19,Fuji23,Fuji25}, although FOPTs remain widely used in phenomenological EOS models.

In this work, we investigate the possibility and implications of FOPTs in NSs by analyzing the coupled nature of gravity and microscopic interactions in the Tolman--Oppenheimer--Volkoff (TOV) equations\,\cite{TOV39-1,TOV39-2,Misner1973}, taking into account basic physical requirements such as thermodynamic consistency and mechanical stability. The TOV equations describe the hydrostatic equilibrium of a spherical NS in general relativity (GR), relating the radial pressure gradient to the local energy density and the enclosed mass. By directly investigating the intrinsic properties of the TOV equations, we can assess the physical viability of FOPTs in a model-independent manner, without assuming any specific input EOS. This novel approach allows us to evaluate the feasibility of sharp transitions and to identify limitations inherent to FOPT-based NS EOSs. In particular, we emphasize that successful numerical integrations of the TOV equations alone do not guarantee the physical consistency of the underlying EOSs.

This work is complementary to existing research on FOPTs in NSs, e.g., see the relevant sections of the reviews\,\cite{Shapiro1983,Oertel17,Baym18,Isa18,Bai19,Dri21,Lovato22,Soren2023,Chat24,Alar25}. Through a direct examination of the intrinsic structure of the TOV equations, we aim to provide new insights that may contribute to a more comprehensive understanding of possible FOPTs in NSs. In particular, our approach reveals physics aspects that may not be immediately evident in the input EOS-based conventional solutions of the TOV equations, thereby enriching previous results and are useful for developing more self-consistent treatments of phase transitions in NSs.

The remainder of the paper is organized as follows. In Section \ref{SEC_2}, we present the physical requirements indicating that a FOPT is disfavored in NSs. Section \ref{SEC_3} demonstrates that a successful integration of the TOV equations does not necessarily reveal a physically inconsistent EOS. Concise and illustrative numerical examples are given in Section \ref{SEC_4}; and finally, Section \ref{SEC_5} presents our conclusions and outlook.

\section{Physical Requirements Forbidding a FOPT in NSs}\label{SEC_2}

The structure of a NS is governed by the TOV equations, written in dimensionless form as\,\cite{CLZ23-a,CLZ23-b,CL24-a,CL24-b,CL25-a,CL25-b}
\begin{equation}\label{def-TOV}
\frac{\d\widehat{P}}{\d\widehat{r}}=-\frac{\widehat{M}\widehat{\varepsilon}}{\widehat{r}^2}\left(1+\frac{\widehat{P}}{\widehat{\varepsilon}}\right)\left(1+\frac{\widehat{r}^3\widehat{P}}{\widehat{M}}\right)\left(1-\frac{2\widehat{M}}{\widehat{r}}\right)^{-1},~~\frac{\d\widehat{M}}{\d\widehat{r}}=\widehat{\varepsilon}\widehat{r}^2.
\end{equation}
Here, $\widehat{P}=P/\varepsilon_{\mathrm{c}}$, $\widehat{\varepsilon}=\varepsilon/\varepsilon_{\mathrm{c}}$, 
$\widehat{M}=M/Q$, and $\widehat{r}=r/Q$ are dimensionless variables; the scale $Q=1/\sqrt{4\pi\varepsilon_{\mathrm{c}}}$ defines the characteristic length (or mass) in units with $G=c=1$, 
and $\varepsilon_{\mathrm{c}}$ is the NS central energy density.

In NSs, the TOV equations couple the pressure, energy density, and enclosed mass in a highly nonlinear and nontrivial manner. This strong interdependence distinguishes NSs from conventional thermodynamic systems, such as liquids or gases, where properties are typically governed by local thermodynamic relations rather than by global gravitational constraints. In an NS, mathematical properties of these quantities, such as continuity and smoothness, influence the system coherently and shape the global structure and stability of the star. 
Consequently, any phase structure in dense matter that emerges under the constraints of hydrostatic equilibrium and gravity may exhibit behaviors different from those in ordinary systems\,\cite{Siem83,Pana84,YGMa97,YGMa99,Elli02,Nat02,Lop05,YGMa05,XuJ13,Deng16,LiBA18,Liu19,RWang20,CaoYT23,He23,He23-a,JHChen24,Deng22,Deng24,Liu22,MaYG23,Boeh22}.

From Eqs.\,(\ref{def-TOV}), the derivative $\d\widehat{P}/\d\widehat{r}$ can vanishe only at two special points: 
the center ($\widehat{r}=0$), where $\widehat{M}(\widehat{r})\approx\widehat{r}^3/3+\cdots$\,\cite{CLZ23-a}, 
and the surface, where the energy density $\widehat{\varepsilon}\to0$. 
At all interior points $0<\widehat{r}<\widehat{R}$, the right-hand side is strictly negative because $\phi=\widehat{P}/\widehat{\varepsilon}=P/\varepsilon>0$, 
$\widehat{r}^3\widehat{P}/\widehat{M}>0$, and $2\widehat{M}/\widehat{r}<1$. 
Thus we obtain:
\begin{equation}\label{PR}
\boxed{
\text{physical constraint:}~~
\frac{\d\widehat{P}}{\d\widehat{r}} < 0,~~\rm{for}~~
0 < \widehat{r} < \widehat{R}.
}
\end{equation}
where $\widehat{R}=R/Q$ defined as the vanishing point of $\widehat{P}$ is the dimensionless NS radius\,\cite{CL25-a}. 

Condition (\ref{PR}) establishes that the pressure inside a NS must be a 
strictly decreasing function of radial coordinate.
Any physically consistent EOS should preserve this monotonicity throughout the NS.

If a FOPT were to occur inside a NS, the pressure profile as a function of radial coordinate would inevitably develop a plateau, since
\begin{equation}\label{def-Ps2}
\boxed{
\frac{\d\widehat{P}}{\d\widehat{r}}=\frac{\d\widehat{P}}{\d\widehat{\varepsilon}}\frac{\d\widehat{\varepsilon}}{\d\widehat{r}}=s^2\frac{\d\widehat{\varepsilon}}{\d\widehat{r}}.}
\end{equation}
The reasoning proceeds in two steps: (1) $\widehat{\varepsilon}(\widehat{r})$ is continuous owing to basic physical requirements; and (2) consequently $\d\widehat{P}/\d\widehat{r}=0$ follows from $s^{2}=0$.

Physically, the energy density $\widehat{\varepsilon}(\widehat{r})$ should vary continuously with the radial distance $\widehat{r}$; in other words, it cannot undergo an abrupt jump within the NS at a given radial coordinate. This is because matter cannot sustain an infinite change in energy density over an infinitesimally thin shell.
Moreover, a discontinuity in $\widehat{\varepsilon}(\widehat{r})$ would unavoidably produce a discontinuity in $\mathrm{d}\widehat{P}/\mathrm{d}\widehat{r}$ through the TOV equations, since $\d \hP/\d\widehat{r}\sim(\widehat{\varepsilon}+\widehat{P})$. This is physically unacceptable, since hydrostatic equilibrium requires a smooth pressure gradient to continuously counteract gravity at every $\widehat{r}$. In the Newtonian limit, for example, the local gravitational force density $-\widehat{M}\widehat{\varepsilon}/\widehat{r}^2$ remains continuous; GR could not change the qualitative feature of this relation (i.e., the force density is still continuous).
As a result, $\widehat{\varepsilon}(\widehat{r})$ remains continuous (although it may not be necessarily smooth) throughout the NS, and the derivative $\d \widehat{\varepsilon}/\d \widehat{r}$ is correspondingly finite.
Since the NS stellar structure equations impose $\mathrm{d}\widehat{P}/\mathrm{d}\widehat{r} \sim (\widehat{\varepsilon} + \widehat{P})$, this situation differs qualitatively from a conventional liquid-gas phase transition, which occurs in spatially uniform systems without gravitational stratification or a preferred direction.

\renewcommand*\figurename{\small FIG.}
\begin{figure}[h!]
\centering
\includegraphics[width=8.cm]{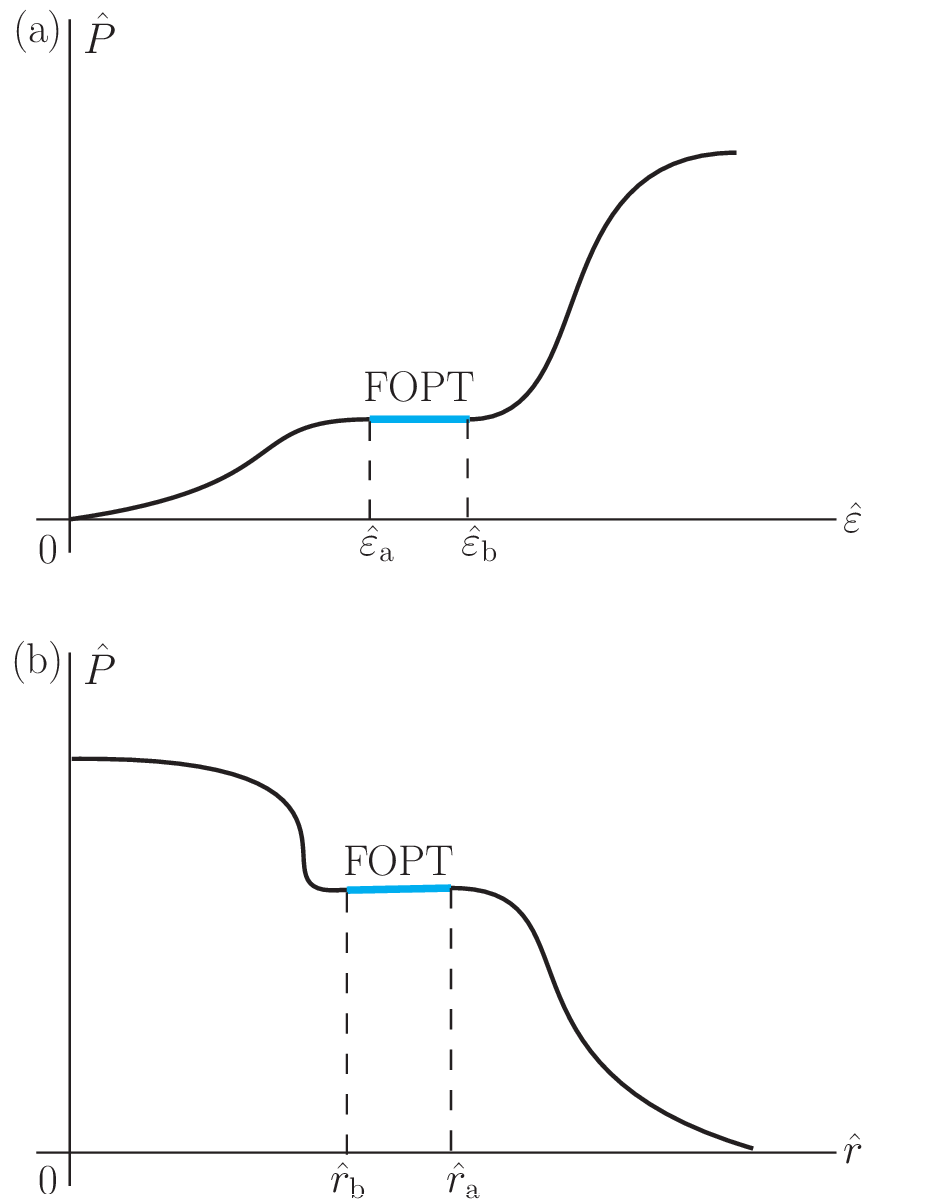}
\caption{(Color Online).  Sketch of a FOPT: a plateau in the EOS correspondingly produces a plateau in the pressure as a function of radial coordinate.
}\label{fig_FOPT}
\end{figure}
In addition, based on the fundamental thermodynamic relation known as the Hugenholtz--Van Hove theorem, $\varepsilon + P = \mu \rho$\,\cite{HVH1958}, which is a very general result in many-body physics relating the energy density $\varepsilon$, chemical potential $\mu$, baryon density $\rho$ and pressure $P$ in any interacting system at zero temperature and is independent of the specific interaction details or model, we have
\begin{equation}
\widehat{\varepsilon} + \widehat{P} = \widehat{\mu} \widehat{\rho}, ~~
\widehat{\mu} = {\mu \rho_{\rm c}}/{\varepsilon_{\rm c}}, ~~
\widehat{\rho} = {\rho}/{\rho_{\rm c}},
\end{equation}
where $\widehat{\mu}$ and $\widehat{\rho}$ are the reduced chemical potential and baryon density, reduced in terms of $\varepsilon_{\rm c}/\rho_{\rm c}$ and $\rho_{\rm c}$, respectively, with $\rho_{\rm c}$ denoting the central baryon density.
In hydrostatic equilibrium, both $\widehat{P}(\widehat{r})$ and $\widehat{\mu}(\widehat{r})$ vary continuously with $\widehat{r}$, as a discontinuity would imply infinite forces or unphysical particle fluxes.  
Similarly, the microscopic baryon density $\widehat{\rho}(\widehat{r})$ cannot jump abruptly in a stable NS (how could some intermediate density region exist otherwise?).
It follows that the energy density $\widehat{\varepsilon}(\widehat{r})$ is also continuous throughout.
Specifically, the HVH theorem can be recast as\,\cite{HVH1958}:
\begin{equation}\label{HV-int}
    \hP=\hrho^2\frac{\partial}{\partial\hrho}\left(\frac{\heps}{\hrho}\right)\to\heps(\hr)=\heps(\hrho(\hr))=\hrho\int^{\hrho}\d\hrho'\left(\frac{\hP(\hrho')}{\hrho'^2}\right)+\rm{const.},
\end{equation}
so the continuity of $\hP(\hr)=\hP(\hrho(\hr))$ naturally enforces the continuity of $\heps(\hr)$, even though their derivatives may not necessarily be smooth.
In the simplest approximation $\varepsilon(\rho) \approx m_{\rm N} \rho$ or $\heps(\hrho)\approx(m_{\rm N}\rho_{\rm c}/\varepsilon_{\rm c})\hrho$ with $m_{\rm N}$ the nucleon mass, the continuity of $\hrho(\hr)$ enforces that of $\heps(\hr)$; however in this case $\hP=0$ is inconsistent with reality.

Furthermore, from the viewpoint of quantum field theory\,\cite{Peskin1995}, the energy density $\langle \mathcal{T}^{00}\rangle$ obtained from the energy-momentum tensor $\mathcal{T}^{\mu\nu}$ is a local function of the underlying quantum fields $\psi$ and their derivatives $\partial_\mu\psi$, which obey the smooth Euler--Lagrange equation; the finite-range particle interactions prevent abrupt changes.
In a FOPT, the transition occurs across a finite microscopic interface, so the energy density interpolates continuously between phases within the star.
The continuity of the energy density implies that $\widehat{M}(\hr)$ is a smooth function of the radial coordinate $\hr$.

These discussions coherently indicate that, within a NS, both the pressure $\hP(\hr)$ and the energy density $\heps(\hr)$ are continuous functions of the radial coordinate $\hr$.
A discontinuity in $\heps(\hr)$, where the energy density immediately inside a point $\hr$ is $\heps_{\rm{left}}$ and immediately outside is $\heps_{\rm{right}}$, would imply that the stellar matter ``skips'' all intermediate local equilibrium states between $\heps_{\rm{left}}$ and $\heps_{\rm{right}}$, which is incompatible with a continuous self-gravitating fluid governed by finite-range interactions.

Consequently, Eq.\,(\ref{def-Ps2}) implies that $\d\widehat{P}/\d\widehat{r}=0$ within the region $\widehat{r}_{\rm b}\leq\widehat{r}\leq\widehat{r}_{\rm a}$, if $s^2=0$ (the hallmark of a FOPT) for $\widehat{\varepsilon}_{\rm a}\leq\widehat{\varepsilon}\leq\widehat{\varepsilon}_{\rm b}$. A sketch of this plateau on the pressure-radial distance curve is shown in FIG.\,\ref{fig_FOPT}. Accordingly,
\begin{equation}\label{EOS-FOPT}
\boxed{
\mbox{EOS with a FOPT}\leftrightarrow\frac{\d\widehat{P}}{\d\widehat{r}}=0,~~\text{for}~~\widehat{r}_{\rm b}\leq\widehat{r}\leq\widehat{r}_{\rm a}.}
\end{equation}
If $\widehat{r}_{\rm a} = \widehat{r}_{\rm b}$, then $\d\widehat{P}/\d\widehat{r} = 0$ occurs only at a single point, so that the region where $s^2 = 0$ is effectively a point (i.e., the FOPT is in “contact”); under such circumstances, the presence of a FOPT necessarily has a negligible impact on the NS structure.

Since the relation (\ref{EOS-FOPT}) arises from a model assumption, whereas the condition (\ref{PR}) represents a fundamental physical requirement derived from the intrinsic structure of the TOV equations, which couples gravity and microscopic interactions, any inconsistency between them must be resolved in favor of the latter. Therefore, the physical constraint (\ref{PR}) formally excludes the occurrence of a FOPT inside NSs.

Practically, when an EOS containing one or more FOPTs is employed in integrating the TOV equations, nothing apparently unusual arises--particularly in the resulting NS M-R relation. This may partly explain why such EOSs are often adopted without much scrutiny, as long as their predictions are consistent with observational constraints. Interestingly, it is precisely the underlying physical requirement (Eq.\,(\ref{PR})) that ensures this ``apparent consistency'' during the integration. In Section \ref{SEC_4}, we shall demonstrate that the TOV integration itself cannot effectively reveal a physically inconsistent EOS, which may nevertheless produce an M-R curve that appears reasonable with respect to NS observations. This subtle yet crucial point calls for a careful re-examination of FOPTs employed in NS modeling.

\section{TOV Integrations Cannot Detect Unphysical EOSs}\label{SEC_3}

Numerical integrations of the TOV equations may fail to reveal an unphysical behavior even when the EOS contains a plateau.
Suppose the pressure is constant over a finite region, $\widehat{P}= \rm{const.}$,  for $\heps_{\rm a}\leq\heps\leq\heps_{\rm b}$ or for $\widehat{r}_{\rm b} \leq \widehat{r} \leq \widehat{r}_{\rm a}$, as shown in FIG.\,\ref{fig_FOPT}. Under the simplest Euler integration scheme, the next-step pressure is
\begin{align}
\widehat{P}_{n} \equiv &
\widehat{P}(\widehat{r}_n)
= \widehat{P}(\widehat{r}_{\rm{b}}+\Delta\widehat{r})\notag\\
\approx& \widehat{P}_{\rm b} -\Delta \widehat{r}
\overbrace{\left[\frac{\widehat{M}\widehat{\varepsilon}}{\widehat{r}^2}
\left(1+\frac{\widehat{P}}{\widehat{\varepsilon}}\right)
\left(1+\frac{\widehat{r}^3 \widehat{P}}{\widehat{M}}\right)
\left(1-\frac{2\widehat{M}}{\widehat{r}}\right)^{-1} \right]_{\rm b}}^{\mbox{positive-definite for $0<\widehat{r}<\widehat{R}$}}.
\end{align}
By construction, $\widehat{P}_n < \widehat{P}_{\rm b}$, i.e., the numerical integration always enforces a decreasing pressure. Similarly, the TOV equations ensure that the next-step pressure 
\begin{equation}
\widehat{P}_{n'}=\widehat{P}(\widehat{r}_n+\Delta\widehat{r})<\hP_n,
\end{equation} etc.; therefore the plateau in the input EOS is not reproduced and no obvious anomaly appears, although some apparent features may emerge in the M-R relation (see Section \ref{SEC_4} for examples). This occurs because the TOV equations intrinsically encode the physical requirement $\d\widehat{P}/\d\widehat{r}<0$ as shown in the basic relation (\ref{PR}), which dominates the numerical evolution. Advanced integration methods, such as Runge-Kutta, cannot overcome this built-in bias: the integration always produces a continuously decreasing pressure profile, though it is not necessarily smooth, regardless of whether the input EOS used is consistent or not with the fundamental physics requirements.

As a result, features of a FOPT may not manifest in the numerical integration, yet this does not guarantee that the underlying EOS is physically sound. In other words, TOV solutions can create an illusion of consistency even with an EOS physically inconsistent with the fundamental requirements. Based on a similar argument, even using an EOS whose slope $\d\widehat{P}/\d\widehat{\varepsilon}$ in some regions is negative, the TOV integration can still enforce a monotonically decreasing pressure, see Section \ref{SEC_4} for numerical examples.

We provide an alternative demonstration for the requirement of the absence of a FOPT in NSs by examining the derivatives of the pressure with respect to the radial coordinate.  For generality, we consider a linear EOS holding within some energy density region:\begin{equation}\label{def-linearEOS}
\widehat{P}(\heps) = f \widehat{\varepsilon} + \mathcal{C}, \quad f\text{ and }\mathcal{C} = \text{const., and }0\leq f\leq1,
\end{equation}
and investigate whether any inconsistency arises. A straightforward derivation yields
\begin{align}\label{def-2P2r}
&\frac{\d^2\widehat{P}}{\d\widehat{r}^2}
=\frac{\widehat{\varepsilon}+\widehat{P}}{f\widehat{r}^4}
\left(1-\frac{2\widehat{M}}{\widehat{r}}\right)^{-2}
\times\Big[(2f+1)\widehat{r}^6\widehat{P}^2+(7f+2)\widehat{r}^3\widehat{M}\notag\\
+&(1-f)\widehat{M}^2+f\widehat{r}\widehat{M}\left(2+\widehat{r}^2\widehat{\varepsilon}\right)-f\widehat{r}^4\widehat{\varepsilon}-f\widehat{r}\widehat{P}\left(1+\widehat{r}^2\widehat{\varepsilon}\right)
\Big],
\end{align}
where we have used the relation $
{\d\widehat{\varepsilon}}/{\d\widehat{r}}=f^{-1}{\d\widehat{P}}/{\d\widehat{r}}$
and $\d\widehat{M}/\d\widehat{r}=\widehat{r}^2\widehat{\varepsilon}$.
Eq.\,(\ref{def-2P2r}) is generally valid for the linear EOS (\ref{def-linearEOS}) and does not rely on  additional assumptions.

For a FOPT ($f=0$) at arbitrary $0<\widehat{r}<\widehat{R}$, we then have
\begin{equation}
\frac{\d^2\widehat{P}}{\d\widehat{r}^2}=\frac{\widehat{M}^2\widehat{\varepsilon}}{f\widehat{r}^4}\overbrace{
\left(1+\frac{\widehat{P}}{\widehat{\varepsilon}}\right)\left(1+\frac{\widehat{r}^3\widehat{P}}{\widehat{M}}\right)^2\left(1-\frac{2\widehat{M}}{\widehat{r}}\right)^{-2}}^{\mbox{finite and positive-definite}}\to\infty,
\end{equation}
as well as
\begin{equation}
\frac{\d^3\widehat{P}}{\d\widehat{r}^3}=\frac{\widehat{M}^3\widehat{\varepsilon}}{f^2\widehat{r}^6}\overbrace{
\left(1+\frac{\widehat{P}}{\widehat{\varepsilon}}\right)\left(1+\frac{\widehat{r}^3\widehat{P}}{\widehat{M}}\right)^3\left(1-\frac{2\widehat{M}}{\widehat{r}}\right)^{-3}}^{\mbox{finite and positive-definite}}\to\infty,
\end{equation}
etc., which are manifestly unphysical: under the assumption of a constant pressure, all derivatives should vanish and cannot diverge.

Moreover, according to the HVH theorem expressed in the form (\ref{HV-int}), the pressure and energy density corresponding to the model (\ref{def-linearEOS}) can be obtained as functions of \(\hrho\) as follows:
\begin{equation}\label{lPPP}
    \hP(\hrho)=fB\hrho^{f+1}+\frac{\mathcal{C}}{f+1},~~\heps(\hrho)=B\hrho^{f+1}-\frac{\mathcal{C}}{f+1},
\end{equation}
as well as $\widehat{\mu}(\hrho)={\partial\heps}/{\partial\hrho}=(f+1)B\hrho^f$, where $B>0$ is the integration constant. Setting $f = 0$ gives $\hP = \mathcal{C}$ and $\heps = B \hrho - \mathcal{C}$, so that the energy density increases while the pressure remains constant. While this behavior can occur in ordinary homogeneous systems, it cannot happen in a NS: the radial structure requires the pressure to decrease outward and the baryon density to be continuous (so that $\heps(\hr)$ is continuous). In particular, the continuity of $\heps(\hr)$ or $\hrho(\hr)$ would produce a plateau in the $\hP(\hr)$ profile, which is inconsistent with the TOV equations (see Eq.\,(\ref{EOS-FOPT})). This demonstrates that the radial structure alone disfavors any EOS segment where the pressure is constant while the energy density jumps.
Similarly for $f=1/3$, we have from Eqs.\,(\ref{lPPP}) that $\hP(\hrho)=B\hrho^{4/3}/3+3\mathcal{C}/4$ and $\heps(\hrho)=B\hrho^{4/3}-3\mathcal{C}/4$, here $-3\mathcal{C}/4$ is the bag constant in the MIT-bag model\,\cite{Chod74}.
On the other hand, for an ``ultra-stiff'' EOS with \( f = 1 \):
\begin{equation}
\hP(\hrho) = B\hrho^2 + \frac{\mathcal{C}}{2}, \quad
\heps(\hrho) = B\hrho^2 - \frac{\mathcal{C}}{2},
\end{equation}
which requires $\mathcal{C} \le 0$ since physically $\hP \le \heps$ to be casual.
Therefore, at very high densities, we have $\hP \approx \heps \sim \hrho^2$,
corresponding to the Zel’dovich model\,\cite{Zeld61}, which considers a vector-field
interaction. This model is characterized by $s^2 \to 1$ and $\hP / \heps \to 1$ in the ultra-dense limit.
For both the MIT‑bag and Zel’dovich models with $f \neq 0$, the behavior remains regular and no singularities arise.

This simple demonstration confirms that numerical integration of the TOV equations cannot detect an unphysical EOS with a FOPT, even though such an EOS is fundamentally inconsistent with the TOV equations themselves.
Instead, one must examine the TOV equations in their differential form to reveal such intrinsic inconsistencies, as these may otherwise be washed out during the integration process.
Mathematically, even if a function is discontinuous at certain points, its integral can be continuous, e.g., 
$\int^{\hr}_0\d\hr\hr^2\heps$ is continuous even if $\heps(\hr)$ is not (unless $\heps(\hr)$ has a Dirac-$\delta$ singularity).

\section{Model Illustrations}\label{SEC_4}

In this section, we present several model calculations of the M-R relation for EOSs that include a FOPT. Although the resulting M-R curves appear self-consistent, their features originate directly from the FOPT and may reflect underlying inconsistencies. Moreover, when the EOS contains inconsistent segments with negative SSS, the M-R curves can still appear reasonable and behave very similarly to those obtained with FOPTs or positive SSS. These results underscore the need for cautious physical interpretation of a FOPT in NS matter.

\begin{figure*}
\centering
\includegraphics[height=7.5cm]{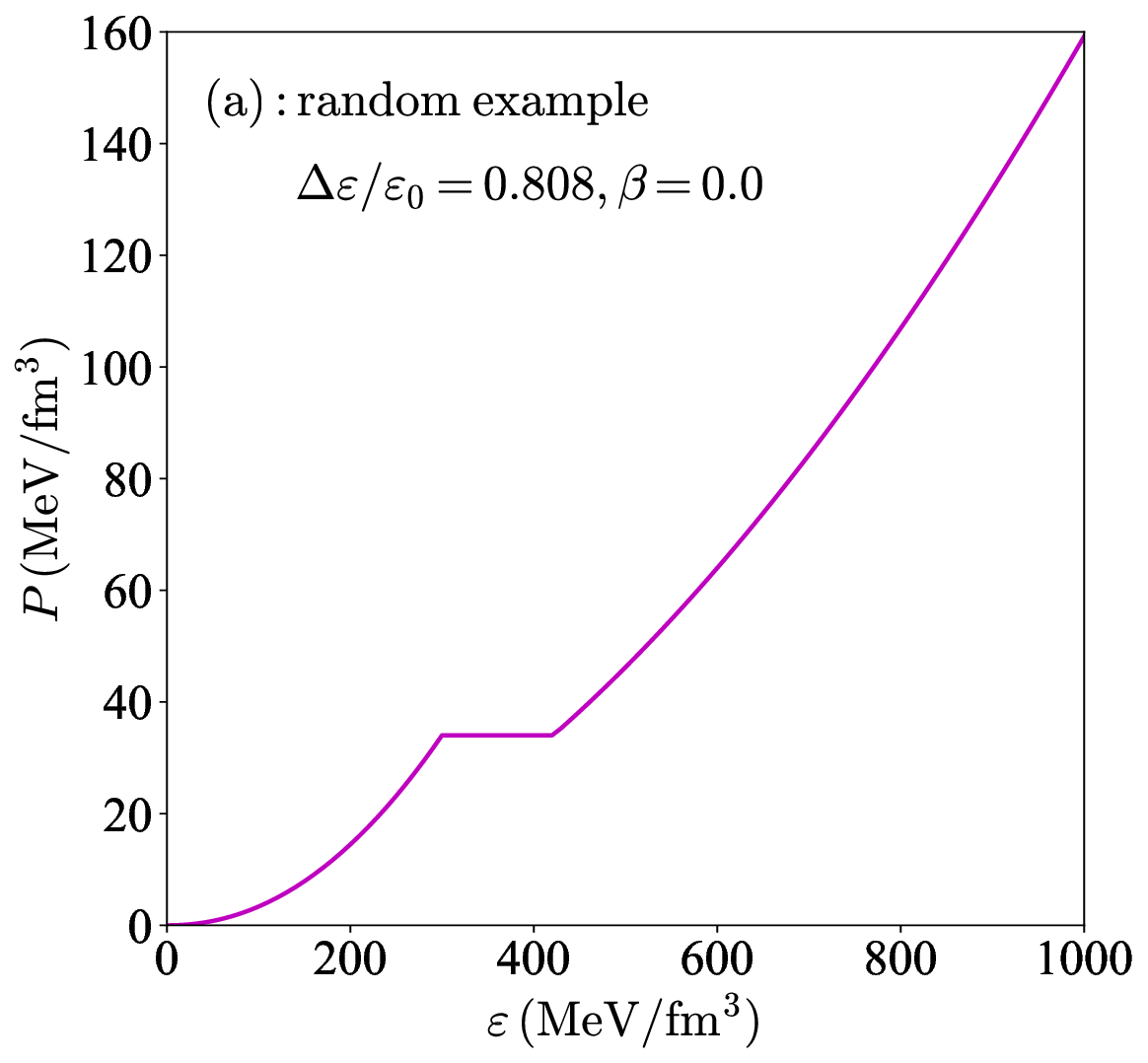}\quad
\includegraphics[height=7.5cm]{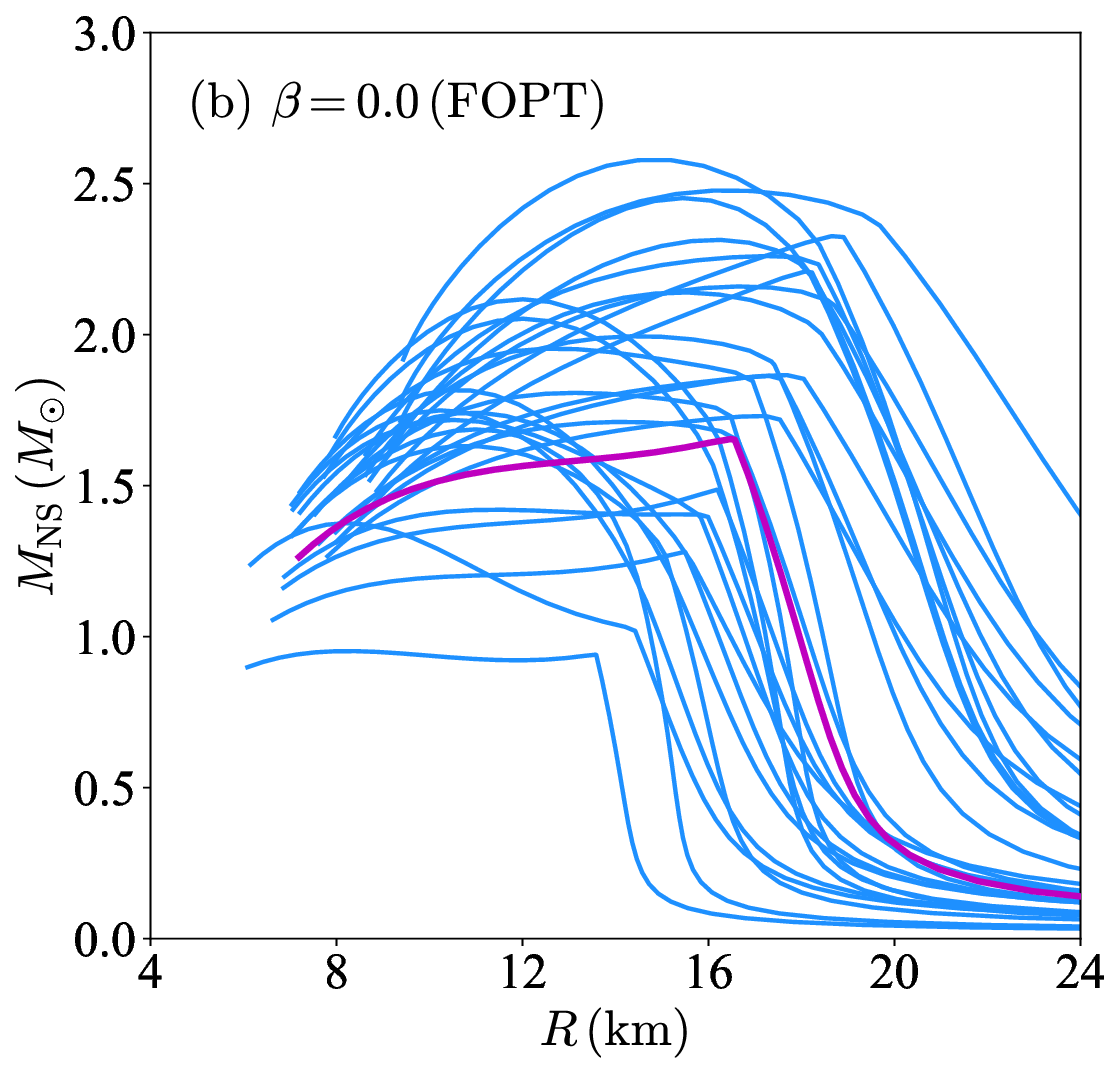}\\
\includegraphics[height=7.5cm]{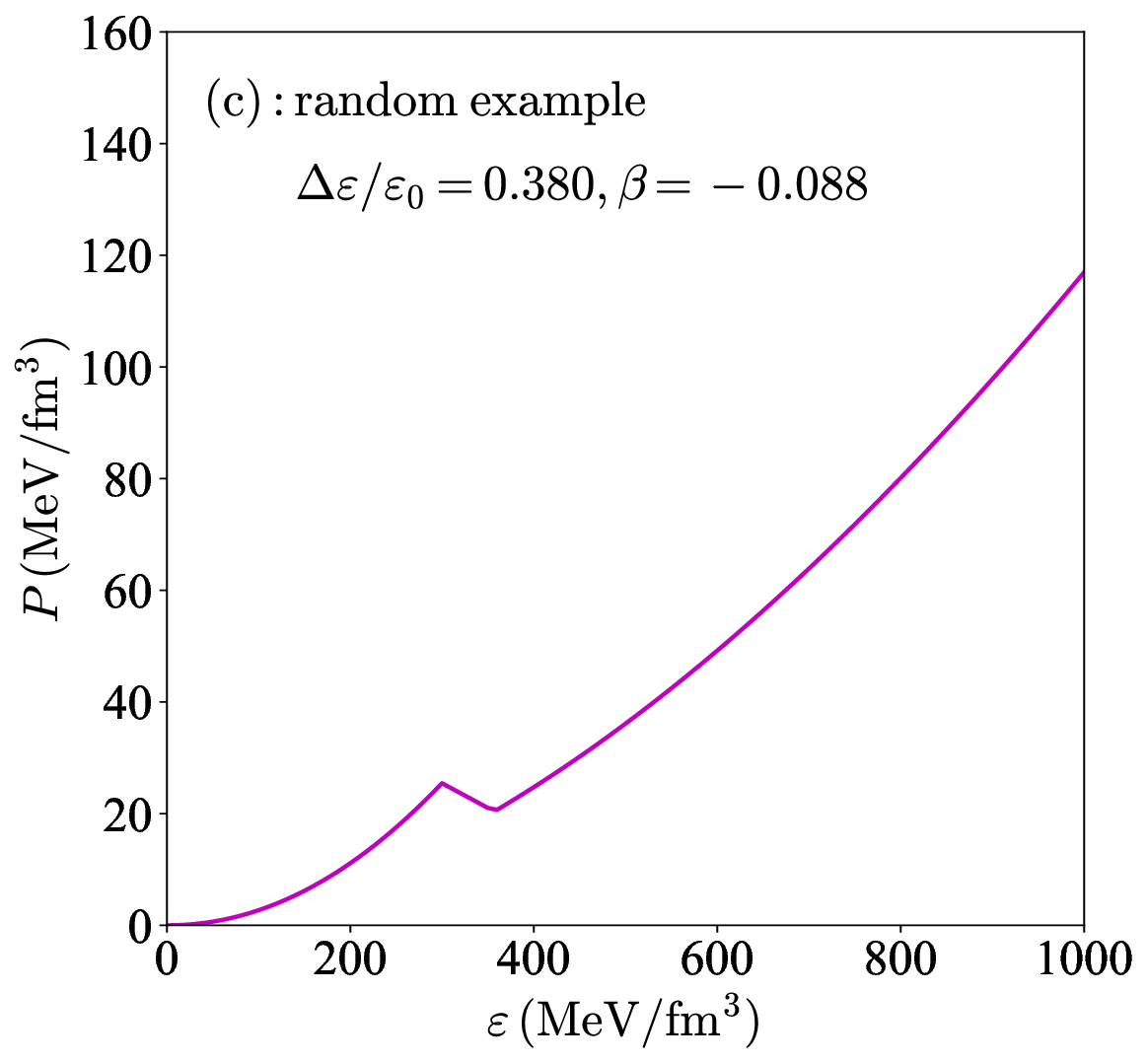}\quad
\includegraphics[height=7.5cm]{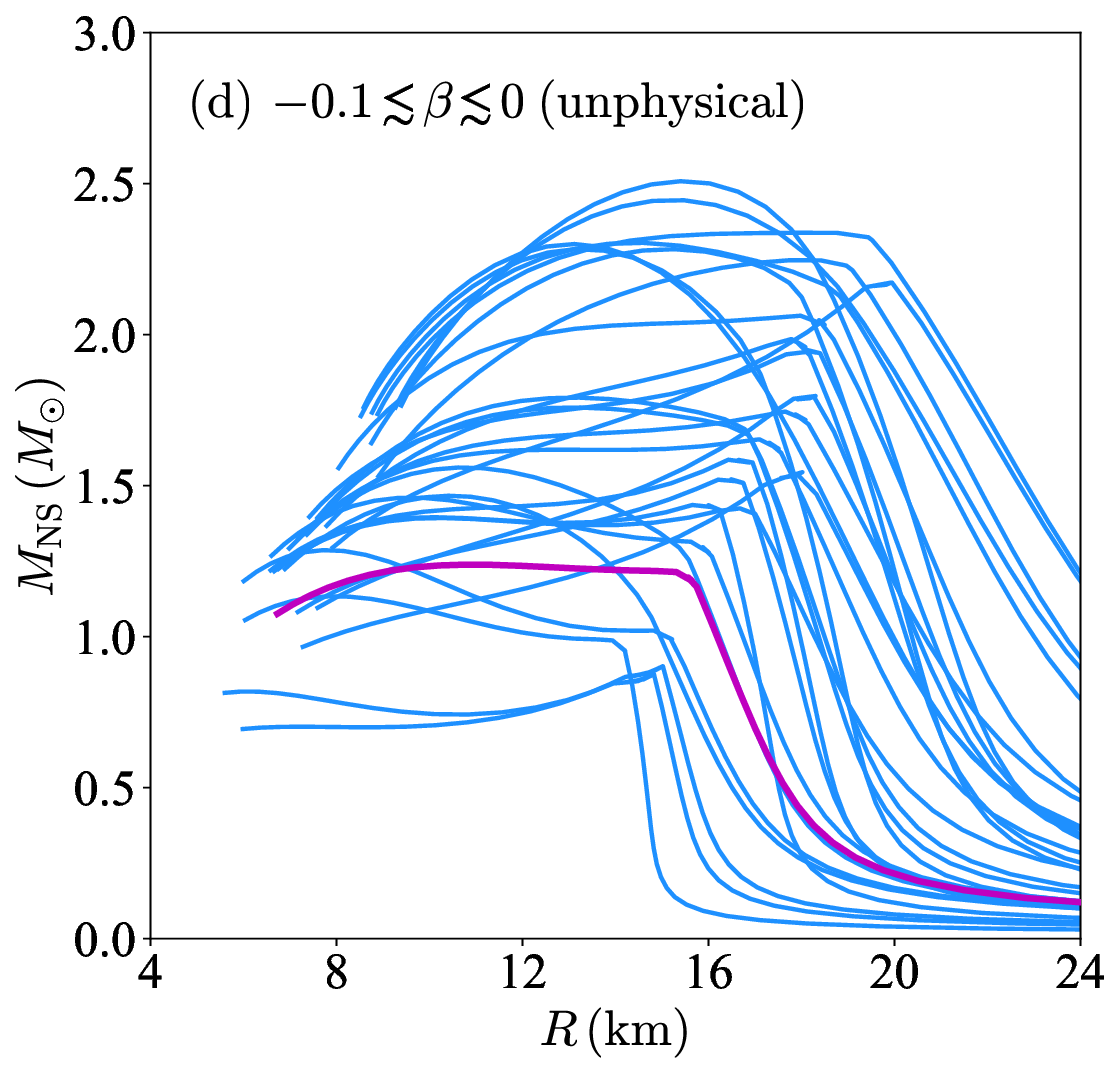}
\caption{(Color Online). First row: NS M-R curves with a FOPT (panel (b)), where $\Delta\varepsilon = \varepsilon_{\rm b} - \varepsilon_{\rm a}$ is randomly selected between 0 and $\varepsilon_0$. Panel (a) shows an example EOS, and the magenta line in panel (b) corresponds to the EOS shown in panel (a).
Second row (panels (c) and (d)): Same as the first row, but allowing a negative effective SSS parameter in the range of $-0.1 \lesssim \beta \lesssim 0$.
}\label{fig_M-R_case}
\end{figure*}

To this end, we employ the following piecewise EOS:
\begin{equation}\label{def-polyEOS}
P(\varepsilon)=\left\{\begin{array}{ll}
\mathcal{K}_{\rm{a}}\varepsilon^{\gamma_{\rm{a}}},&0\leq\varepsilon\leq\varepsilon_{\rm a},\\
P_{\rm{a}}+\beta\left(\varepsilon-\varepsilon_{\rm{a}}\right),&\varepsilon_{\rm a}\leq\varepsilon\leq\varepsilon_{\rm b},\\
\mathcal{K}_{\rm b}\varepsilon^{\gamma_{\rm b}},&\varepsilon\geq\varepsilon_{\rm b},
\end{array}\right.
\end{equation}
where $\mathcal{K}_{\rm a} = P_{\rm a}/\varepsilon_{\rm a}^{\gamma_{\rm a}}$, and $\varepsilon_{\rm a}/\varepsilon_0 = 2$ with $\varepsilon_0 \approx 150\,\rm{MeV}/\rm{fm}^3$ denoting the energy density at nuclear saturation.
For illustration, we adopt $P_{\rm a} \approx 20$-$50\,\rm{MeV}/\rm{fm}^3$\,\cite{LCXZ21} (randomly selected), and the polytropic indices are taken as $\gamma_{\rm a} \approx 2$-$13/6$ and $\gamma_{\rm b} \approx 5/3$-$2$, respectively. The intermediate linear segment, characterized by the slope $\beta$ (effective SSS parameter), connects the two polytropes.
Physically, $\beta$ should be non-negative (and it plays the role of $f$ in Eq.\,(\ref{EOS-FOPT}) if  $0\leq\beta\leq1$); however, we will show that even a negative $\beta$ can yield seemingly reasonable M-R relations. The pressure at the upper boundary of the intermediate region is given by
$P_{\rm b} = P_{\rm a} + \beta(\varepsilon_{\rm b} - \varepsilon_{\rm a})$,
where $\varepsilon_{\rm b}/\varepsilon_0 \geq 2$. If $\varepsilon_{\rm b}/\varepsilon_0 = 2$, the effective parameter $\beta$ has no effect since $\Delta\varepsilon = \varepsilon_{\rm b} - \varepsilon_{\rm a} = 0$. Furthermore, if $\varepsilon_{\rm b}/\varepsilon_0 = 2$ and $\gamma_{\rm a} = \gamma_{\rm b}$, the EOS reduces to a single, smooth polytropic form. The coefficient $\mathcal{K}_{\rm b}$ is determined by $\mathcal{K}_{\rm b} = P_{\rm b}/\varepsilon_{\rm b}^{\gamma_{\rm b}}$.
The SSS is given by 
\begin{equation}
s^2 ={\d P}/{\d\varepsilon}=
\begin{cases}
\phi_{\rm a}\gamma_{\rm a} ({\varepsilon}/{\varepsilon_{\rm a}} )^{\gamma_{\rm a}-1}, & 0 \le \varepsilon \le \varepsilon_{\rm a},\\
\beta, & \varepsilon_{\rm a} \le \varepsilon \le \varepsilon_{\rm b},\\
\phi_{\rm b}\gamma_{\rm b}
( {\varepsilon}/{\varepsilon_{\rm b}})^{\gamma_{\rm b}-1}, & \varepsilon \ge \varepsilon_{\rm b},
\end{cases}
\end{equation}
where $\phi_{\rm a}=P_{\rm a}/\varepsilon_{\rm a}$ and $\phi_{\rm b}=P_{\rm b}/\varepsilon_{\rm b}$.
Under $\beta=0$ (FOPT), the $s^2$ is discontinuous at both $\varepsilon_{\rm a}$ and $\varepsilon_{\rm b}$, jumping from $\phi_{\rm a}\gamma_{\rm a}$ to $0$ at $\varepsilon_{\rm a}$ and from $0$ to $\phi_{\rm b}\gamma_{\rm b}$ at $\varepsilon_{\rm b}$.
One naturally has $\phi_{\rm b} < \phi_{\rm a}$ (under $\beta=0$) if $\Delta \varepsilon \neq 0$ since now $\varepsilon_{\rm b}>\varepsilon_{\rm a}$ and $P_{\rm a}=P_{\rm b}$, which is fundamentally inconsistent with the intrinsic constraint that the EOS-parameter $\phi = P/\varepsilon$ decreases monotonically outward from the center when the radial coordinate is small\,\cite{CLM25-a}, further supporting the view that a FOPT is disfavored near the NS center from another perspective. In fact, the constraint $\phi_{\rm b}\ge\phi_{\rm a}$ requires $\beta\ge \phi_{\rm a}$.
Nonetheless, this minimal construction\,\cite{Shapiro1983} is sufficient for our purpose to capture and illustrate the essential physical impact of the $\beta$ parameter on NS M-R relation.

Shown in panel (b) of FIG.\,\ref{fig_M-R_case} are the M-R curves for $\beta = 0$, corresponding to a FOPT, where $0\lesssim\Delta\varepsilon=\varepsilon_{\rm b}-\varepsilon_{\rm a}\lesssim\varepsilon_0$ is randomly generated. As an example of the EOSs generated, the EOS corresponding to the magenta line is shown in panel (a). It is seen that some M-R curves exhibit a change in the sign of the slope $\d M_{\rm{NS}}/\d R$, switching from negative for large-radius NSs to positive for small-radius ones (the latter being unstable), while others show only a change in the slope magnitude without sign reversal, which remain negative. In all cases, a discontinuity in the slope $\d M_{\rm{NS}}/\d R$ is present, reflecting the FOPT feature of the input EOSs.
These behaviors can be understood from the Seidov condition, which states that a stable hybrid-star branch remains connected to the ordinary NS branch if the energy-density discontinuity $\Delta\varepsilon$ at the transition is smaller than a critical threshold $\Delta\varepsilon_{\rm{crit}}/\varepsilon_{\rm a}=(1+3\phi_{\rm a})/2$\,\cite{Seidov1971,Lind98,AHP13}.

Although the Seidov condition\,\cite{Seidov1971,Lind98} is widely used to assess the stability of NSs containing a FOPT, it only constrains the global hydrostatic response to an energy density discontinuity. Therefore, it does not ensure the internal consistency of the EOS itself, such as the thermodynamic continuity or local/microscopic mechanical stability (e.g., the gravitational force cannot change abruptly). In fact, derived purely as a macroscopic stability criterion, it implicitly assumes that an EOS with a density discontinuity can be inserted consistently. In particular, it does not verify whether the EOS satisfies deeper thermodynamic or causal consistency conditions, including continuity of chemical potential and pressure across the transition, monotonicity and continuity of the energy density and pressure profiles ($\d\widehat{P}/\d\widehat{r} < 0$, $\d\widehat{\varepsilon}/\d\widehat{r} < 0$, with both $\widehat{P}$ and $\widehat{\varepsilon}$ continuous along the radial coordinate), and local mechanical stability. Consequently, EOSs that are intrinsically inconsistent with the physical constraint of Eq. (\ref{PR}) can still yield seemingly reasonable M-R curves by numerically integrating the TOV equations (which can effectively wash out the inconsistencies as discussed earlier), highlighting the need for independent physical consistency checks.

To further illustrate our main points, we now consider model EOSs containing an unphysical segment with a negative effective SSS parameter $\beta$, as shown in panel (c) of FIG.\,\ref{fig_M-R_case} for one example; here $-0.1 \lesssim \beta \lesssim 0$ is chosen arbitrarily. Despite this apparent pathology, numerical integration of the TOV equations can still be performed, producing the M-R relation shown in panel (d) of FIG.\,\ref{fig_M-R_case}. Remarkably, the resulting M-R curves appear reasonable at first glance, even though the underlying EOS is internally inconsistent with the fundamental physics requirements.

\section{Summary and Outlook}\label{SEC_5}

In summary, by examining the coupled nature of gravity and the strong interaction within the TOV equations, together with basic requirements of thermodynamic consistency and mechanical stability, we have shown that a FOPT in NSs is generally disfavored. Numerical TOV integrations can yield apparently reasonable M-R relations even for unphysical EOSs, underscoring the importance of independent checks for thermodynamic consistency. From a modeling perspective, EOSs incorporating FOPTs may serve as approximate or limiting schemes. If the plateau width $\Delta\varepsilon/\varepsilon_0$ is very small or collapses to a single point, the FOPT exerts little influence on the NS structure, as indicated by the Seidov condition. In particular, the ``contact'' FOPT with $\Delta\varepsilon/\varepsilon_0 = 0$, corresponding to an inflection point on the $P$-$\varepsilon$ curve, is practically undetectable. Idealized plateaus within NSs cannot capture the full microscopic physics or metastable states that arise from complex particle interactions. 

Our analysis indicates that a genuine FOPT would violate the continuity and differentiability conditions essential for hydrostatic equilibrium and thermodynamic consistency. Specifically, a continuous fluid cannot sustain a discontinuous energy density, as such a jump would violate the covariant energy-momentum conservation, $\nabla_\mu \mathcal{T}^{\mu\nu} = 0$. To treat such a discontinuity consistently, one must introduce a thin shell stress-energy layer at the transition interface, with the Israel junction conditions\,\cite{Israel66,Pois04,Pere15,Dong25} specifying how the metric adjusts across the boundary.

While phenomenological FOPT-based EOSs remain useful as limiting approximations for exploring extreme scenarios such as twin-star configurations, continuous crossover EOSs provide a more physically consistent and theoretically grounded description of dense stellar matter. They remain smooth over the entire density range, avoiding the energy-density discontinuities characteristic of FOPTs. Lattice quantum chromodynamics (QCD) simulations at finite temperature and chemical potential indicate that the hadron-quark transition proceeds as a smooth crossover\,\cite{Aoki06,Baza12,Bhat14,Burg18,Tani20}. Extending this behavior to cold, dense matter suggests that continuous EOSs not only avoid unphysical energy-density jumps but also align with first-principles expectations of the fundamental strong interaction. Crossover EOSs require no special treatment at phase boundaries, yield smoother thermodynamic and stellar profiles, and naturally satisfy key continuity constraints on the chemical potential, pressure, and energy density.

Fundamentally, the TOV equations are intrinsically composition-degenerate in the sense that, as long as the same EOS is employed, regardless of how it is constructed, the same M-R relation will result. It is therefore challenging to identify observable signatures of phase transitions in NSs. From an observational standpoint, current measurements of NS masses and radii remain too limited and imprecise to clearly discriminate between M-R relations arising from a FOPT and those from a smooth EOS.

Looking ahead, as larger and more precise observational datasets become available, it should become possible to determine which classes of EOSs most faithfully capture the true behavior of dense matter in NS interiors, thereby providing stringent constraints on the EOS and guiding the development of physically consistent NS models. For instance, by assuming a non-negative dimensionless trace anomaly $\Delta = 1/3 - \hP/\heps \geq 0$, Ref.\,\cite{Fuji22} constrained the M-R relation to approximately satisfy $R/\rm{km}\gtrsim 4.84M_{\rm{NS}}/M_{\odot}$.
As observational precision continues to improve, one may directly test whether the measured M-R relations follow the predictions of FOPT-based EOSs or the bound cited here. Future data may also allow discrimination among alternative EOS assumptions and model frameworks. In this context, it is particularly encouraging that forthcoming high-precision joint measurements of NS mass and radius, especially from next-generation X-ray timing missions\,\cite{Wat16,Pap20,LiA25,NewA,Yunes22} and gravitational-wave observations\,\cite{Yunes22,GW1,GW2,Magg20,Bran23,Gitt24}, could enable direct probes of the smoothness, slope and curvature of the NS M-R relation. Such observations may serve as indirect tests of whether the dense-matter EOS contains any discontinuous segment.

\section*{Acknowledgement} We would like to thank Xavier Grundler, Rui Wang, Wen-Jie Xie, Nai-Bo Zhang and Zhen Zhang for helpful discussions. This work was supported in part by the National Natural Science Foundation
of China under contract No. 12147101, the U.S. Department of Energy, Office of Science, under Award Number DE-SC0013702, the CUSTIPEN (China-U.S. Theory Institute for Physics with
Exotic Nuclei) under the US Department of Energy Grant No. DE-SC0009971.

\section*{Data Availability}
The data that support the findings of this article will be openly available\,\cite{Data}.

\end{document}